\documentclass[structabstract]{aa}
\usepackage{txfonts}
\usepackage{graphicx}
\usepackage{natbib}
\bibpunct{(}{)}{;}{a}{}{,}

\def\dsct{$\delta$~Scuti}
\def\ds{$\delta$~Scuti}

\def\deg{$^{\circ}$}

\def\Msun{$M_{\odot}$}

\def\Ha{H$_{\alpha}$}
\def\vsini{$v\cdot\sin{i}$ }

\def\k1c{$\kappa^1$ Ceti}

\def\freqs{frequencies}
\def\freq{frequency}

\def\lc{light curve}

\def\Tf{$T_{\rm{eff}}$}
\def\lg{log\,$g$}
\def\hd{HD\,61199}
\def\vsini{$v\sin i$}
\def\kms{$\rm{kms^{-1}}$}
\def\cd{$d^{-1}$}

\begin{document}

\title{MOST\thanks{Based on data from the MOST satellite, a Canadian Space Agency mission, jointly operated by Dynacon Inc., the University of Toronto Institute for Aerospace Studies and the University of British Columbia, with the assistance of the University of Vienna and on spectra taken with the Coud\'e-echelle spectrograph attached to the 2-m telescope of the Thueringer Landessternwarte Tautenburg.} discovers a multimode \ds\ star in a triple system: HD\,61199}
\author{M. Hareter \inst{1}\thanks{\email{hareter@astro.univie.ac.at}}
\and O. Kochukhov \inst{2}
\and H. Lehmann \inst{3}
\and V. Tsymbal \inst{1,4}
\and D. Huber \inst{1}
\and P. Lenz \inst{1}
\and W. W. Weiss \inst{1}
\and J. M. Matthews \inst{5}
\and S. Rucinski \inst{6}
\and J. F. Rowe \inst{6}
\and R. Kuschnig \inst{1}
\and D. B. Guenther \inst{7}
\and A. F. J. Moffat \inst{8}
\and D. Sasselov \inst{9}
\and G. A. H. Walker \inst{6}
\and A. Scholtz \inst{10} 
}
\institute{Institut f\"ur Astronomie, T\"urkenschanzstrasse 17, 1180 Vienna, Austria
\and Department of Physics and Astronomy, Uppsala University, Box 515, 751 20 Uppsala, Sweden
\and Th\"uringer Landessternwarte Tautenburg, 07778 Tautenburg, Germany 
\and Tavrian National University, Dep. Astronomy, Simferopol, Ukraine
\and Department of Physics \& Astronomy, University of British Columbia, 6224 Agricultural Road, Vancouver, B.C., V6T 1Z1, Canada
\and David Dunlap Observatory, Department of Astronomy, University of Toronto, Toronto, Ontario, L4C 4Y6, Canada
\and Department of Astronomy and Physics, St. Mary's University, Halifax, Nova Scotia, NS B3H 3C3, Canada
\and D\'epartement de physique, Universit\'e de Montr\'eal, Montr\'eal, Qu\'ebec, QC H3C 3J7, Canada
\and Harvard-Smithsonian Center for Astrophysics, Cambridge, Massachusetts, MA 02138, USA
\and University of Technology, Institute of Communications and Radio-Frequency Engineering, Gusshausstrasse 25/389, 1040 Vienna, Austria
}

\abstract
{A field star, \hd\ (V $\approx$ 8), simultaneously observed with Procyon by the MOST (Microvariability \& Oscillations of STars) satellite in continuous runs of 34, 17, and 34 days in 2004, 2005, and 2007, was found to pulsate in 11 frequencies in the \ds\ range with amplitudes from 1.7 down to 0.09 mmag.  The photometry 
also showed variations with a period of about four days.  To investigate the nature of the longer period, 45 days of time-resolved spectroscopy was obtained at the Th\"uringer Landessternwarte Tautenburg in 2004.  The radial velocity measurements indicate that \hd\ is a triple system.}
{A \ds\ pulsator with a rich eigenspectrum in a multiple system is promising for asteroseismology. Our objectives were to identify which of the stars in the system is the \ds\ variable and to obtain the orbital elements of the system and the fundamental parameters of the individual components, which  are constrained by the pulsation frequencies of the \ds\ star.}
{Classical Fourier techniques and least-squares multi-sinusoidal fits were applied to the MOST photometry to identify the pulsation frequencies. The groundbased spectroscopy was analysed with least-squares-deconvolution (LSD) techniques, and the orbital elements derived with the KOREL and ORBITX routines. Asteroseismic models were also generated.}
{The photometric and spectroscopic data are compatible with a triple system consisting of a close binary with an orbital period of 3.57 days and a \ds\ companion (\hd\,A) as the most luminous component. The \ds\ star is a rapid rotator with about \vsini = 130\,\kms\  and an upper mass limit of about 2.1\,\Msun. For the close binary components, we find they are of nearly equal mass, with lower mass limits of about 0.7\,\Msun. Comparisons to synthetic spectra indicate these stars have a late-F spectral type. The observed oscillation frequencies are compared to pulsation models to further constrain the evolutionary state and mass of \hd\, A. The orbit frequency of the close binary corresponds to the difference of the two {\dsct} frequencies with the highest amplitudes -- a coincidence that is remarkable, but not explained.}
{}


\maketitle

\section{Introduction}
\label{sect-intro}

The \ds\ variables are stars of intermediate mass located in the classical instability strip near the main sequence, which pulsate in radial and nonradial acoustic (p-) modes driven by opacity effects; see review by \citet{Breger99}. Asteroseismic modelling of \ds\ stars has been a challenge, even in cases where dozens of
pulsation frequencies have been identified -- e.g., FG Vir \citet{Breger05} and HD\,209775 \citet{Matthews07} - because the theoretical eigenspectra are so densely populated with frequencies. Constraining the fundamental parameters of a \ds\ star in a multiple system through its orbital elements is one possible avenue to a 
unique asteroseismic model of the star.

\hd\ (V = 7.97) was observed by MOST, a Canadian space telescope dedicated to high-precision photometry of bright stars, in 2004, because it happened to lie in the field of MOST's first primary science target, Procyon \citep{Matthewsetal04}.  Little was known about \hd\ beyond its HD Catalogue classification as an 
A3 star. \citet{Gatewood} give a parallax of 5.2$\pm$0.9\,mas and likely luminosity class of V (see their Table 2, in which \hd\ corresponds to AO 1098). The 2004 MOST photometry showed clear oscillations with periods in the \ds\ range, plus an obvious variation with a periodicity near 4 days.  If the latter was due to rotational modulation, it would have given the rotation period of the star and helped for more clearly interpreting and modelling any rotational splitting in the pulsation spectrum.  Because of this possibility, groundbased spectroscopy was obtained in 2004 from Th\"uringer Landessternwarte Tautenburg to investigate the \hd\ system further.  The star was again monitored by MOST in 2005 and 2007 as part of subsequent Procyon photometric runs.

Photometric indices are available for \hd, which can be employed to obtain an initial estimation of the stellar parameters. We have used TempLoggTNG \citep{KupkaBruntt01,Kaiser06} and published Str\"omgren indices $b-y = 0.125, m_1 = 0.192, c_1 = 0.931$ and $ H_{\beta}  = 2.816$ \citep{HauckM} for various calibrations to calculate values of {\Tf} ranging from $7664\,\rm{K}$ to $7811\,\rm{K}$ and {\lg} from 3.94 to 3.80 (see Table 1), based on different calibrations \citep{Moon,Ribas,Balona84,Napi}. These values were derived from the combined light of the triple system and imply a spectral type of A7 or A8, significantly later than the HD classification. The uncertainties in the individual temperature determinations are at least 200 K, and in the surface gravities, at least 0.3\,dex. 

\begin{table}

\caption{ \Tf\ and \lg\ of \hd\ derived from published Str\"omgren colours.}

\begin{tabular}{lcccc}

\hline\hline
                    & Moon & Ribas et al. & Balona & Napiwotzki et al. \\
\hline
{\Tf}              & 7811   & 7843          & 7745    & 7664 \\
{\lg}              & 3.94    & 3.94           & 3.95     & 3.80 \\
spectral type & A7      & A7             & A7       & A8 \\
\hline

\end{tabular}

\label{tab-calib}
\end{table}

\hd\ is kinematically similar to DG Leo \citep{Fremat}, where two Am stars form a binary system and a third more distant component is a \ds\ star. This system is bright enough to decide which star is pulsating by means of radial velocity variations, what would need larger telescopes for \hd.

The MOST photometry and data reduction is presented in Section\,\ref{sect-Mph}, followed by the frequency analysis in Section\,\ref{sect-FreqAna}. The spectroscopy and their analyses are described in Section\,\ref{sect-spectro}, and the resulting fundamental parameters and orbital solution are presented. Asteroseismology is attempted in Section\,\ref{sect-Model}, where the observed pulsation frequencies are compared to models.

\section{MOST Photometry and Data Reduction}
\label{sect-Mph}

The Canadian MOST (Microvariability \& Oscillations of STars) microsatellite houses a 15-cm telescope feeding a CCD photometer through a single broadband optical filter.  For descriptions of the MOST mission before and after launch, see \citet{Walker03} and \citet{Matthews04}. MOST's Sun--synchronous 820-km polar orbit (period = 101.4 min) allows uninterrupted observations of a target located in the satellite's continuous viewing zone for up to 8 weeks. MOST now can collect, simultaneously, three types of photometric data: (a) Fabry Imaging -- in which the telescope entrance pupil is imaged as an annulus (diameter ~ 40 pixels), for very bright stars ("Fabry Targets"), minimizing the influence of satellite pointing jitter and cosmic ray strikes on the photometry, and enhancing the total signal \citep{Matthews04}. (b) Direct Imaging -- in which a slightly defocused image of a star (a "Direct Imaging Target") is projected directly onto the CCD \citep{Rowe}. (c) Guide Star photometry -- in which photometry of up to about 30 stars used for MOST's startracking attitude control system is processed on board and downlinked to Earth.  Since MOST's launch in June 2003 and its commissioning, the MOST team has improved the pointing performance of the satellite, improving the Direct Imaging photometric performance.  The 
Guide Star photometry mode was implemented during the mission as well, but was not available during the 2004 observations of \hd\ presented here.
 
\hd\ was observed simultaneously with Procyon from Jan. 8 to Feb. 9,  2004 and from Jan. 24 to Feb. 10, 2005 as a Direct Imaging Target, and from Jan. 4 to Feb. 12, 2007 as a Guide Star Target.  The integration times 
were about 1 second in the first two runs, while the effective integration time was 24.8 seconds in the 2007 run (in which short exposures were stacked on board). Table \ref{tab-details} lists details for the photometry.  The pointing performance and other aspects of MOST photometry during the 2007 Procyon were the best achieved to that point in the mission, resulting in significantly reduced point-to-point scatter.

\begin{table} 
 \caption{Details of the MOST photometry of \hd.}
 \label{tab-details}
 \centering
 \begin{tabular}{l c c c}
    
    \hline \hline
    Year & 2004 & 2005 & 2007\\
    \hline
    
    Length of dataset [d] & 33.57 & 16.97 & 34.25\\
    Integration time [s] & 0.9 & 1 & 24.8 \\
    Sampling time [s] & 6 & 5 & 37 \\
    \# of raw data-points  & 108\,912 & 49\,978 & 78\,176\\
    \# of data-points after reduction & 89\,592 & 47\,079 & 60\,249\\
    point-to-point scatter [mmag] & 4.9 & 4.2 & 2.6\\
        
    \hline
  \end{tabular}
 
\end{table}

For the Direct Imaging photometry of \hd\ in 2004 and 2005, a reduction routine developed by \citet{Huber} was used, similar to the decorrelation technique developed by \citet{Reegen06} for MOST Fabry Imaging, but which also takes into account satellite pointing errors.  Both techniques decorrelate the signal present in the target pixels with those in the background pixels.  The 2007 Guide Star photometry of \hd\ is handled in a similar way, but for these data, there is no direct information about the background sky in each Guide Star subraster.  The on-board Guide Star data processing is performed as follows: A Guide Star is sampled within a 20 $\times$ 20 pixel subraster. The subrasters are read out roughly every second and the average of the first and last columns of the subraster are used for an on-board background correction. This procedure is repeated until the final integration time of the Primary Science Target in the field is reached.  Then the individual background-corrected pixel intensities are integrated. Only this sum and the Guide Star exposure time are stored in the FITS header of the Primary Science Target data file which is downloaded. Residual stray light due to scattered Earthshine remains in the Guide Star light curves sent to Earth. Fortunately, the modulation of the stray light with the MOST orbital period is well known, and 15 Guide Stars were observed simultaneously with \hd\, allowing for a modified decorrelation technique using non-variable stars in the field to correct for residual instrumental and 
background variations.  

\begin{figure} 
 \resizebox{\hsize}{!} {
    \includegraphics*{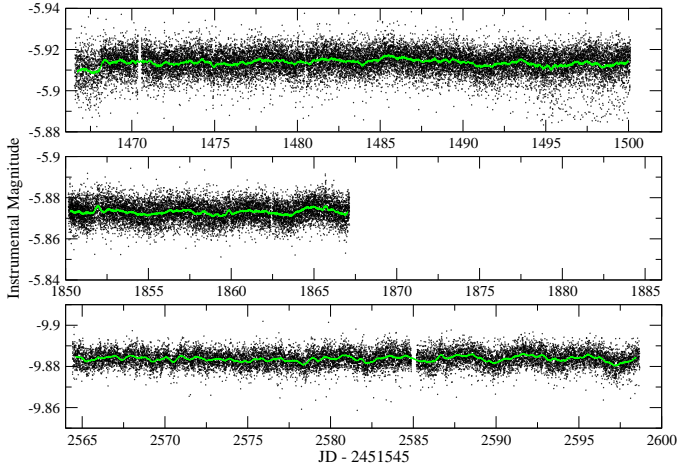}
  }
 \caption{MOST light curves of \hd\ from 2004 (top panel), 2005, and 2007 (bottom panel). The solid line shows a running average of 600 data points.}
 \label{fig-lcs}
\end{figure}

\begin{figure}
 \resizebox{\hsize}{!} {
    \includegraphics*{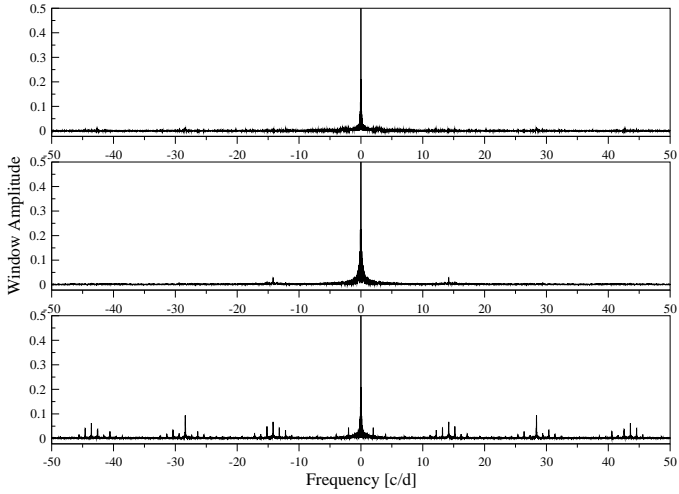}
 }
 \caption{Spectral windows for 2004, 2005 and 2007 from top to bottom. Aliasing is practically not present in the Direct Imaging data of 2004 and 2005. The more restrictive outlier rejection policy in the Guide Star photometry of 2007 affects the spectral window more than the Direct Imaging reduction.}
 \label{fig-spw}
\end{figure}

Figure\,\ref{fig-lcs} shows the reduced MOST light curves of \hd\ for the three observing runs and their respective spectral windows are shown in Fig.\,\ref{fig-spw}. The low-amplitude alias sidelobes visible in the spectral window of the 2007 data, spaced by the MOST satellite orbital frequency (164\,{$\mu$}Hz\,$\approx$\, 14.192\,{\cd}), 
are due to our more conservative rejection of outliers in the Guide Star reduction. This procedure is producing more gaps when stray light is highest at certain phases in the MOST orbit than in the 2004 and 2005 Direct Imaging reductions.

The decorrelation method reduced the amplitudes of orbitally-modulated background variations in the Guide Star data to about 8\% of the raw values. For the final {\lc} of the 2007 run, a second-order trend (possibly due to long-term stellar variability or subtle instrumental effects) was subtracted. The subtraction of this polynomial fit affects neither the identified {\freqs} nor their amplitudes, which were checked by comparing the discrete Fourier transform spectra before and after subtraction.

\section{Frequency Analysis}
\label{sect-FreqAna}

SigSpec \citep{Reegen07} and Period04 \citep{LenzB} were adopted for the {\freq} analysis. SigSpec performs iterative prewhitening of the data with the most significant frequency found in each iteration. The iterations end if a specified threshold is reached; the default value is 5.5 in spectral significance \citep{Reegen07}, corresponding to a signal-to-noise (S/N) ratio in amplitude of about 4. A peak identified with a significance of 5.5 means that a similar peak has a probability of only $10^{-5.5}$ of being produced by white noise (i.e., once in $3.2\times 10^5$ cases).

\begin{figure}
 
 \resizebox{\hsize}{!} {
 \includegraphics*{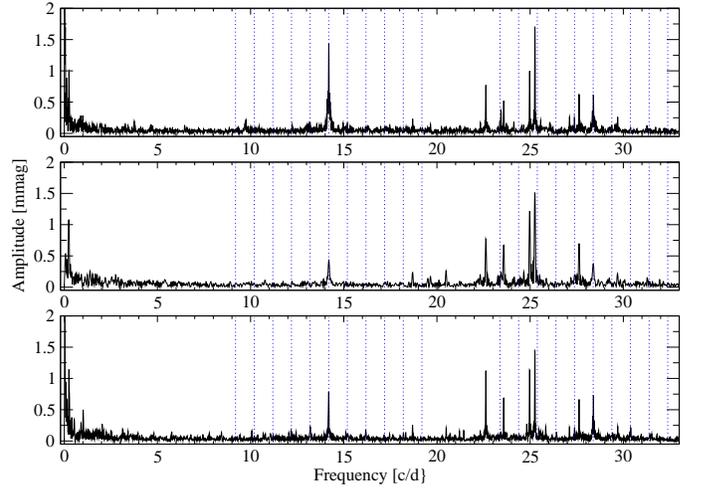}
 }
 \caption{Fourier amplitude spectra of the 2004 (top), 2005 (middle) and 2007 (bottom) light curves. Power due to the stray light modulation with the MOST orbit is present, but with significantly reduced amplitudes compared to the raw data. The vertical dashed lines indicate the MOST orbital frequency, its harmonics and its 1 d$^{-1}$ sidelobes.  (The MOST Sun-synchronous orbit carries it above nearly the same point on Earth after 1 day, resulting in a secondary modulation of the scattered Earthshine.)}
 \label{fig-dft}
\end{figure}

Fig.\,\ref{fig-dft} shows the Fourier amplitude spectra of the individual observing runs. Frequencies were searched in the range $0\, - \,80$\,{\cd}, down to a significance limit of 5.5. The final set of 13 adopted \freqs\ listed in Tab.\,\ref{tab-freqs} includes only those which could be detected in all three runs, where 11 of 
the frequencies are in the {\dsct} domain, and the other two are close to 0.25\,\cd ($f_3$ and $f_7$).  The uncertainties of the frequencies and amplitudes are calculated according to the technique of \citet{Kallinger08}. Based on extensive simulations, those authors find that the frequency errors are overestimated when using the classical Fourier $T^{-1}$ criterion. They propose instead: \\ \\
$\sigma(f) = \frac{1}{T \sqrt{sig(a)} } $ \hspace*{5mm}and\hspace*{4mm} $\sigma(a) = \frac{a}{\sqrt{sig(a)}} $ \\ \\
\noindent
where $\sigma(f)$ and $\sigma(a)$ are the errors in frequency and amplitude, respectively; $T$ is the time base of the observations in days; and $sig(a)$ denotes the significance of an individual frequency with amplitude $a$. 

\begin{table*}
 \caption{List of identified frequencies and amplitudes in the three MOST data sets.}
  \centering
  \begin{tabular}{lrrrrrrrrr}
    
    \hline \hline
    label & frequency & amplitude & signif. & frequency & amplitude & signif. & frequency & amplitude & signif.\\
          & 2004 [\cd]& 2004 [mmag] & 2004 & 2005 [\cd] & 2005 [mmag] & 2005 & 2007 [\cd] & 2007 [mmag] & 2007 \\
    \hline
    
   $f_{1}$ & 25.257 $\pm 0.001$ & 1.72 $\pm 0.05$ & 872  & 25.254 $\pm 0.003$ & 1.53 $\pm 0.07$ & 534 & 25.256 $\pm 0.001$ & 1.46 $\pm 0.05$ & 930 \\
   $f_{2}$ & 24.974 $\pm 0.002$ & 0.98 $\pm 0.06$ & 308  & 24.972 $\pm 0.003$ & 1.19 $\pm 0.06$ & 344 & 24.973 $\pm 0.001$ & 1.14 $\pm 0.05$ & 614 \\
   $f_{3}$ & 0.254  $\pm 0.007$ & 0.84  $\pm 0.06$ & 231  & 0.241 $\pm 0.015$ & 1.06 $\pm 0.06$ & 287 & 0.255 $\pm 0.007$ & 0.98 $\pm 0.04$ & 512 \\
   $f_{4}$ & 22.625 $\pm 0.002$ & 0.79 $\pm 0.06$ & 206 & 22.624 $\pm 0.005$ & 0.78 $\pm 0.06$ & 156 & 22.624 $\pm 0.001$ & 1.14 $\pm 0.05$ & 635 \\
   $f_{5}$ & 27.628 $\pm 0.003$ & 0.62 $\pm 0.05$ & 131 & 27.627 $\pm 0.005$ & 0.70 $\pm 0.06$ & 130 & 27.625 $\pm 0.002$ & 0.70 $\pm 0.04$ & 307 \\
   $f_{6}$ & 23.582 $\pm 0.003$ & 0.53 $\pm 0.05$ & 99 & 23.579 $\pm 0.005$ & 0.67 $\pm 0.06$ & 119 & 23.582 $\pm 0.002$ & 0.73 $\pm 0.04$ & 327 \\
   $f_{7}$ & 0.286  $\pm 0.007$ & 0.42  $\pm 0.05$ & 60 & 0.275 $\pm 0.017$ & 0.20 $\pm 0.06$ & 12 & 0.279 $\pm 0.007$ & 0.65 $\pm 0.04$ & 262 \\
   $f_{8}$ & 18.697 $\pm 0.007$ & 0.23 $\pm 0.05$ & 19 & 18.691 $\pm 0.014$ & 0.16 $\pm 0.04$ & 17 & 18.691 $\pm 0.004$ & 0.25 $\pm 0.04$ & 44 \\
   $f_{9}$ & 23.435 $\pm 0.005$ & 0.35 $\pm 0.05$ & 43 & 23.441 $\pm 0.015$ & 0.24 $\pm 0.06$ & 15 & 23.437 $\pm 0.006$ & 0.18 $\pm 0.04$ & 24 \\
   $f_{10}$ & 29.697 $\pm 0.006$ & 0.27 $\pm 0.05$ & 26 & 29.689 $\pm 0.016$ & 0.22 $\pm 0.06$ & 13 & 29.701 $\pm 0.005$ & 0.24 $\pm 0.04$ & 40 \\
   $f_{11}$ & 22.329 $\pm 0.010$ & 0.15 $\pm 0.05$ & 8.6 & 22.329 $\pm 0.023$ & 0.16 $\pm 0.06$ & 6.5 & 22.326 $\pm 0.007$ & 0.16 $\pm 0.04$ & 18 \\
   $f_{12}$ & 23.372 $\pm 0.009$ & 0.18 $\pm 0.05$ & 12 & 23.359 $\pm 0.025$ & 0.14 $\pm 0.06$ & 5.4 & 23.373 $\pm 0.012$ & 0.09 $\pm 0.04$ & 6.4 \\
   $f_{13}$ & 24.128 $\pm 0.009$ & 0.17 $\pm 0.05$ & 11 & 24.158 $\pm 0.026$ & 0.13 $\pm 0.06$ & 5.1 & 24.130 $\pm 0.009$ & 0.12 $\pm 0.04$ & 10 \\
    
  \end{tabular}
  \label{tab-freqs}

\end{table*}

\begin{figure}
 \resizebox{\hsize}{!} {
   \includegraphics*{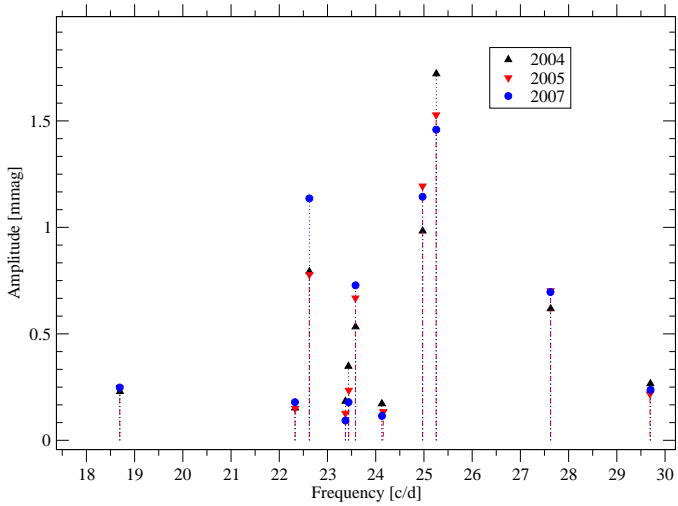}
 }
 \caption{Frequencies in the {\dsct} range detected in {\hd}.  Uncertainties in amplitude (see text) are approximately the sizes of the symbols.}
 \label{fig-ampl}
\end{figure}

The amplitudes $a$ in Table\,\ref{tab-freqs} were calculated from the combined light of the system, which we know to be triple (see Sec.\,\ref{sect-spectro}). Therefore, the intrinsic amplitudes are actually higher by the inverse fraction of light of the pulsating component. We show in Sec.\,\ref{sect-spectro} that about 80\% of the total flux can be attributed to the \ds\ star in the system. Hence, a scaling factor of 1.25 is needed to obtain the intrinsic amplitude. In addition, the pulsation amplitudes do show evidence of variability across the three runs, particularly for $f_4$, as can be seen in Fig.\,\ref{fig-ampl}.  At this stage of the investigation, we can only speculate if this variability is due to unresolved beating frequencies or due to another effect. In a recent study, Breger \& Lenz investigated possible explanations for amplitude variability in the $\delta$ Scuti star 44 Tau \citep{BregerLenz2008}. 

The two close frequencies $f_{9}$ and $f_{12}$ in Fig.\,\ref{fig-ampl} are spaced 0.063 d$^{-1}$ (for 2004 data). The frequency errors given for both frequencies in Table\,\ref{tab-freqs} are 0.005 and 0.009 respectively. Even the more conservative estimation, the $T^{-1}$ criterion gives a frequency resolution of 0.029 d$^{-1}$. Hence the frequencies are resolved. $f_{12}$ is close to an expected sidelobe with order 5 (i.e. $2\times 14.192 - 5\, d^{-1}$) of the first harmonic (28.384 d$^{-1}$) of the MOST orbit frequency. Sidelobes tend to decrease in amplitude with growing order. No sidelobe has been detected below the orbit harmonic of 28.384 d$^{-1}$, except for order 1 at 27.387 d$^{-1}$. Thus, $f_{12}$ has been adopted in the frequency list (Table\,\ref{tab-freqs}).

\begin{figure}
 \resizebox{\hsize}{!} {
   \includegraphics*{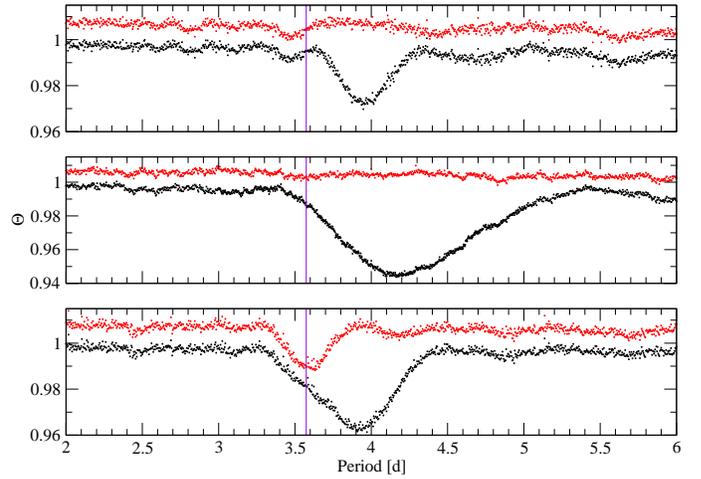}
 }
 \caption{ Phase Dispersion Minimization (PDM) diagrams for the 2004 (top), 2005 (middle) and 2007 (bottom) photometry. The lower curve in each panel is the PDM after prewhitening all significant frequencies higher than 1\,\cd.  Each upper curve shows the PDM of the residuals after prewhitening by the dominant period in the lower curve. The thin vertical line marks the orbital period of the stellar system found spectroscopically (see Sec.\,\ref{sect-spectro}).}
 \label{fig-pdm}
\end{figure}

The original 2005 light curve of \hd\ shows longer-term variability with a period of about 4 days which can be seen by eye (and is not present in the other stars observed in the same field).  This timescale was investigated through Phase Dispersion Minimization \citep[PDM,][]{Stellingwerf78}. The PDM diagrams of the 2004, 2005 and 2007 photometry are plotted in Fig.\,\ref{fig-pdm}.  For this analysis, all significant power at frequencies above 1\,\cd\ has been prewhitened and the data have been averaged in 1-min bins. The most prominent feature in all three PDM diagrams occurs near a period of about 4 days, or more precisely, of 3.94\,$\pm$\,0.12\,days in the 2004 and 2007 data and of 4.15\,$\pm$\,0.24\,days in 2005. For those errors, the error estimation of $0.25T^{-1}$ is adopted, which is given by \citet{Kallinger08} for the case of two closely spaced frequencies. The change in the period is within the errors. Prewhitening of the dominant period in each data set yields the PDM results given in the upper curves in each panel.  This reveals clearly in the 2007 data, the stellar system orbital period determined by spectroscopy (3.57\,d, shown by the vertical line, see Sec.\,\ref{sect-spectro}).  The period may be present marginally in the PDM analysis of 2004 data, but is absent in the 2005 PDM residuals. The SigSpec and Period04 routines find this latter period ($f_7$ in Table\,\ref{tab-freqs}) in all three data sets, within the expected frequency uncertainty. 

\section{Spectroscopy}
\label{sect-spectro}

Prompted by the longer period seen in the 2004 MOST photometry, a spectroscopic follow-up campaign was carried out.  A total of 29 high-resolution spectra ($R$=63\,000, wavelength range from $4760 - 7360$\,{\AA}) spanning 45 days were obtained with the Coud\'e-echelle spectrograph attached to the 2-m telescope of the Th\"uringer Landessternwarte Tautenburg. The Signal to Noise (S/N) of the spectra is between 60 and 180 with a mean value of 130. The S/N was measured in the continuum around 5200\,\AA.

\begin{table}
 \caption{Radial velocities for each component of the \hd\ system derived from the Tautenburg spectra.}
 \label{tab-obslog}
\footnotesize{
\flushright
 \begin{tabular}{r r r r r r }
    \hline \hline
    No. & HJD      & Phase & V$_{Ba}$ & V$_{Bb}$  & V$_ A$\\
        & -2453000 &       & (\kms)& (\kms) & (\kms)\\
    \hline
 01 & 69.2601 & 0.048 & 109.3$\pm$0.3 & -33.3$\pm$0.4 & 46.8$\pm$1.1\\
 02 & 69.2818 & 0.054 & 108.7$\pm$0.3 & -33.0$\pm$0.4 & 47.3$\pm$1.1\\
 03 & 69.3038 & 0.060 & 107.3$\pm$0.4 & -31.3$\pm$0.4 & 48.1$\pm$1.4\\
 04 & 70.2833 & 0.334 & 1.0$\pm$0.3   & 77.6$\pm$0.4 & 45.7$\pm$0.9\\
 05 & 70.3049 & 0.340 & -1.4$\pm$0.3  & 80.3$\pm$0.4 & 46.6$\pm$1.0\\
 06 & 76.3648 & 0.036 & 110.8$\pm$0.7 & -36.6$\pm$1.0 & 44.7$\pm$2.5\\
 07 & 79.3111 & 0.860 & 85.1$\pm$0.4  & -10.3$\pm$0.5 & 45.9$\pm$1.4\\
 08 & 80.3180 & 0.142 & 85.7$\pm$0.5  & -11.4$\pm$0.6 & 45.4$\pm$1.5\\
 09 & 81.2756 & 0.409 & -24.1$\pm$0.3 & 103.6$\pm$0.4 & 45.4$\pm$1.2\\
 10 & 82.2736 & 0.689 & 9.8$\pm$0.3   & 68.0$\pm$0.4 & 43.7$\pm$0.9\\
 11 & 86.3425 & 0.827 & 73.3$\pm$0.4  & 1.8$\pm$0.5 & 46.6$\pm$1.2\\
 12 & 87.3380 & 0.105 & 96.9$\pm$0.4  & -23.3$\pm$0.5 & 46.1$\pm$1.3\\
 13 & 90.2719 & 0.926 & 105.1$\pm$0.6 & -28.8$\pm$0.4 & 45.7$\pm$2.3\\
 14 & 91.3869 & 0.238 & 43.3$\pm$1.0  & 33.7$\pm$1.3 & 45.5$\pm$1.2\\
 15 & 91.4086 & 0.244 & 42.7$\pm$1.3  & 35.1$\pm$1.1 & 46.7$\pm$1.3\\
 16 & 93.3190 & 0.779 & 51.7$\pm$0.4  & 24.5$\pm$0.5 & 45.1$\pm$1.0\\
 17 & 93.3409 & 0.785 & 54.4$\pm$0.3  & 22.1$\pm$0.4 & 44.6$\pm$0.9\\
 18 & 93.4236 & 0.808 & 64.3$\pm$0.3  & 11.0$\pm$0.4 & 45.0$\pm$1.0\\
 19 & 95.3082 & 0.335 & -0.1$\pm$0.3  & 78.1$\pm$0.4 & 46.5$\pm$0.9\\
 20 & 96.3153 & 0.617 & -17.3$\pm$0.3 & 96.0$\pm$0.4 & 47.0$\pm$1.1\\
 21 & 97.3154 & 0.897 & 97.6$\pm$0.3  & -23.8$\pm$0.4 & 46.6$\pm$1.1\\
 22 & 98.3171 & 0.177 & 71.4$\pm$0.4  & 2.6$\pm$0.5 & 44.2$\pm$1.0\\
 23 & 99.3164 & 0.457 & -33.1$\pm$0.3 & 113.0$\pm$0.4 & 46.7$\pm$1.1\\
 24 & 107.3302 & 0.699 & 14.6$\pm$0.5 & 63.5$\pm$0.6 & 45.2$\pm$1.5\\
 25 & 110.2993 & 0.529 & -34.5$\pm$0.3 & 114.2$\pm$0.4 & 46.7$\pm$1.1\\
 26 & 111.2962 & 0.808 & 64.3$\pm$0.3  & 10.7$\pm$0.4 & 44.0$\pm$0.9\\
 27 & 112.3031 & 0.090 & 100.5$\pm$0.4 & -27.7$\pm$0.4 & 46.3$\pm$1.1\\
 28 & 113.2990 & 0.369 & -16.3$\pm$0.4 & 91.6$\pm$0.5 & 45.7$\pm$1.1\\
 29 & 114.3325 & 0.658 & -3.3$\pm$ 0.3 & 80.7$\pm$0.4 & 45.4$\pm$1.0\\
 
    \hline
  \end{tabular}
}
\end{table}

\begin{figure*}
\sidecaption
   \includegraphics*[width=13cm]{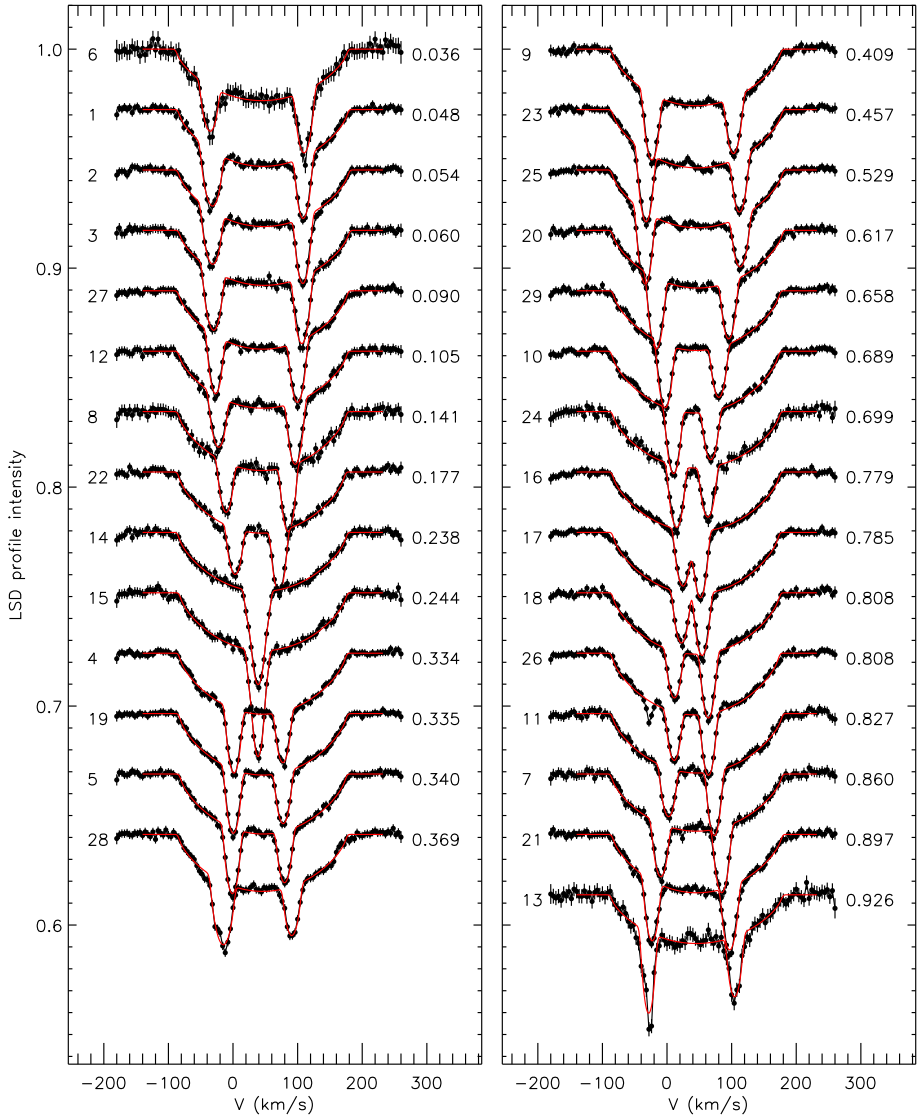}  
 \caption{LSD profiles phased with the final derived orbital period of 3.5744(3) days. The red solid curve shows the fit of the
triple-spectral model. The spectrum numbers (in chronological order) are given on the left of each spectrum, while orbital phase is given on 
the right.}
 \label{fig-lsdprofiles}
\end{figure*}

We analysed the dynamics of the \hd\ system by applying the least-squares deconvolution technique \citep[LSD,][]{Donati97} to the spectra. Starting with preliminary estimates for the atmospheric parameters, we extracted from the Vienna Atomic Line Database \citep[VALD,][]{Kupka99, Ryabchikova97,Piskunov95} 700 spectral lines with a residual intensity higher than 0.2 and used this list for the LSD line mask. Based on Table\,\ref{tab-calib}, following model parameters have been adopted: \Tf\,=\,7800\,K, \lg\,=\,4.0, and solar chemical composition. The relative shape of the LSD profile is rather insensitive to line masks with different parameters. 
Thus, because we are using only relative quantities, no additional error is introduced by the difference of \Tf\ of the components. Fig. \ref{fig-lsdprofiles} shows the LSD profiles of all 29 spectra with a typical S/N of about 500. A broad -- but shallow -- line of a rapidly rotating star (to which we refer as component  A) is clearly present, superposed on the narrow lines of an obvious binary system (referred to as components Ba and Bb). At some phases (e.g., spectrum {\#}26) a fourth faint narrow component appears at the constant velocity of $-26$\,\kms. At the phase of 0.926 (spectrum {\#}13) it blends the lines of one binary component. For the spectra in which the narrow line component is present, the heliocentric velocity of the observer projected in the direction of \hd\ varied in the range of \makebox{-25} to \makebox{-28 \kms}. Hence, this faint component may well be due to stray moonlight. 
The combined light of the system is dominated by the rapidly rotating star, which causes the broad-line component in the LSD profiles. This feature did not change significantly its position and its equivalent line width is much larger than those of the narrow line components. 

We fitted the LSD spectra with a simple model consisting of three superimposed Gaussian profiles, each
broadened with a FWHM\,=\,5\,\kms. The following free model parameters were adjusted using a least-squares procedure: \vsini, radial velocity and intensity. The 
LSD profiles were fitted in a window between $-140$ and $+235$ \kms\ and the resulting best-fit solution is shown in Fig.\,\ref{fig-lsdprofiles}. The 
radial velocities measured from the spectral fits are listed in Table\,\ref{tab-obslog}. Errors of the LSD profiles were estimated from the discrepancy between the LSD model and observations. Uncertainties of the parameters determined from the fit to LSD profiles are formal errors calculated by a least-squares fitting routine.

\subsection{Determination of the Close Binary Orbit}
\label{sect-detorbit}
To determine the orbital elements for the close binary evident in the spectra, two different approaches were used: (i) analysis of the radial velocities determined from the fits to LSD profiles using ORBITX \citep{Tokovinin92} and (ii) disentangling the spectrum with KOREL \citep{Hadrava95}.
  
\begin{figure}
 \resizebox{\hsize}{!} {
   \includegraphics*{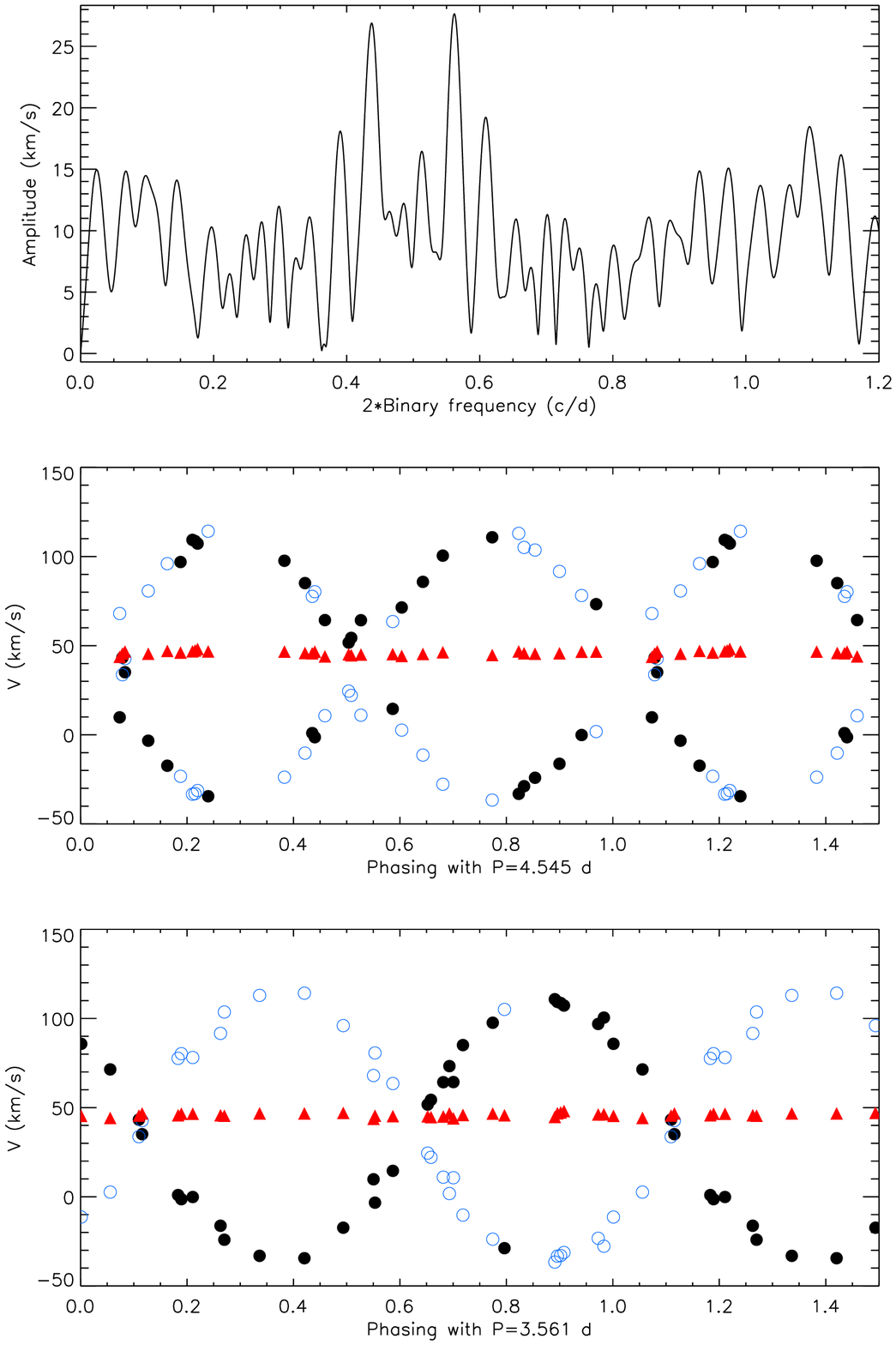}
 }
 \caption{Top panel: Fourier transform of the time-dependent velocity observable, as explained in the text. The two highest peaks correspond to periods of 4.54 and 3.56 days. Note that the radial velocity observable varies with twice the orbit frequency. The middle and bottom panels show the radial velocity curves of the three components of \hd\ phased with 4.54 and 3.56 days respectively. The filled circles correspond to component Ba 
(which has systematically stronger spectral lines, as seen in Figure\,\ref{fig-lsdprofiles} ); open circles correspond to component Bb; and the triangles to the fast rotating component  A.}
 \label{fig-rvdft-rvcurves}
\end{figure}

Due to similar line intensities of the components Ba and Bb it is not straightforward to identify their contribution in the composite line profile. Hence a time-dependent observable, $| V_{\rm{Ba}} - V_{\rm{Bb}} |$, was formed which varies 
with twice the orbital \freq\ of the close binary.  Fig.\,\ref{fig-rvdft-rvcurves} (top panel) shows the resulting Fourier amplitude spectrum. The two highest peaks occur at frequencies of 0.4375 and 0.5616\,\cd, corresponding to  periods of 4.545 and 3.561\,days, respectively. Phase plots with both possible periods are shown in the middle and bottom panels of Fig.\,\ref{fig-rvdft-rvcurves}. 
A period of 4.55\,days leads to discontinuities in the phase diagram, while a period of 
3.56\,days produces a coherent RV curve. The correct identification of components Ba and Bb appears to fail only for spectrum {\#}13 where the the constant feature at \makebox{-26\,\kms\ } interferes, and possibly in spectrum {\#}15 where the LSD profiles of the narrow line components are blended.

Second, KOREL was applied with the following setup: three stars forming two hierarchical orbits, with components Ba and Bb close together and  A in a wide orbit. Spectra of the rapidly rotating B3V star HD\,120315 were used to model the spectrum of telluric lines which were subtracted from the broad-lined component to obtain a pure stellar spectrum.  Orbital elements were made free parameters in the fit and line strengths were allowed to vary with time. Since the line positions of component  A did not change measurably during the period of observations, we assumed the period of the wide orbit to lie in the range $1\,000 - 20\,000$\,d. For 12 different long trial periods, we detected only marginal changes in the disentangled spectra and in the resulting orbital elements. 20\,000\,d is the upper limit to observe any changes in the KOREL solution.

KOREL determines simultaneously orbital elements, line strengths of each component and disentangled spectra. Radial velocities are not determined from the input spectra but are calculated from the orbital solution. 
The KOREL code finds several
local minima that all give zero RMS of the obtained RVs (shifts applied
by KOREL to the individual spectra), i.e. the RVs are identical to those
calculated from the simultaneously obtained orbital elements. We
searched for the best solution nearest to that obtained with ORBITX,
using the wavelength interval 4895\,-\,5680\,\AA. For an error estimation we
then divided this interval into 10 parts of equal lengths and applied
KOREL to each of them. In this way we obtained 10 orbital solutions,
mean and standard deviation, which we weighted by the total equivalent
width included in each subspectrum. There was no significant difference
between the values obtained from the full wavelength interval and the
weighted means.

\begin{figure}
 \resizebox{\hsize}{!} {
   \includegraphics*[angle=-90]{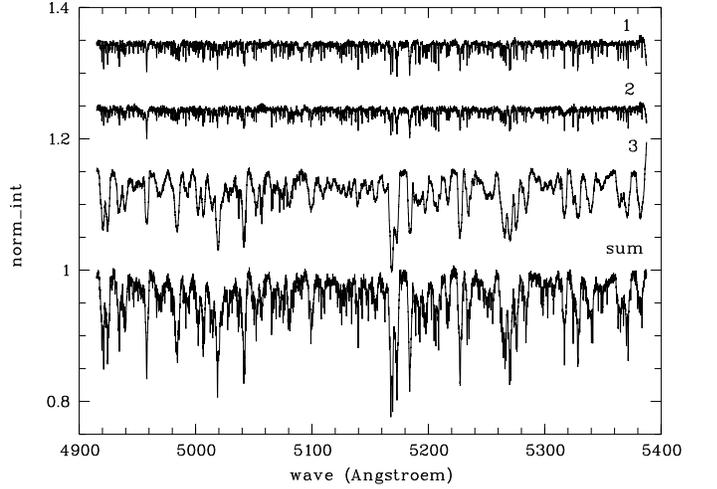}
 }
 \caption{Disentangled spectra of the components Ba (1), Bb (2) , A (3, from top to bottom) and the combined spectrum.}
 \label{fig-spectra}
\end{figure}

The disentangled spectra and the combined spectrum are shown in Fig.\,\ref{fig-spectra}. KOREL provides no information about the individual continua of the stars, because the disentangled spectra are normalised to the common continuum of the observed composite spectrum. To renormalise the KOREL output spectra, one must either use information about the luminosity ratios of the stars or, as we did, scale the stellar spectra by comparison to synthetic spectra.
KOREL has difficulties disentangling the \Ha\ region, where the RV shifts are much smaller than the intrinsic line width. For this reason the disentangled spectra cannot be used for \Tf\ estimation from \Ha.

\begin{figure}
 \resizebox{\hsize}{!} {
   \includegraphics*{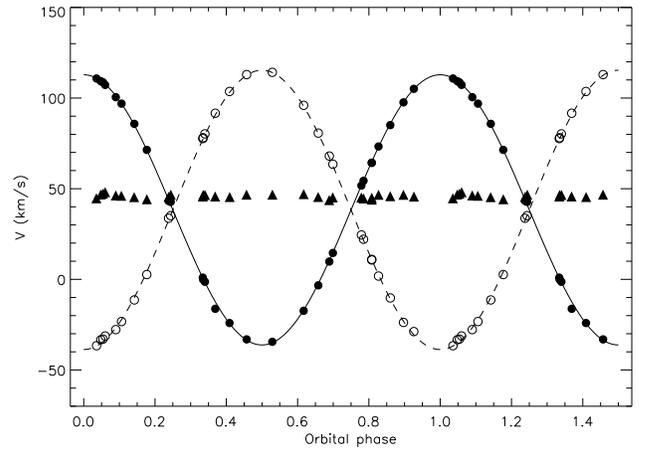}
 }
 \caption{Radial velocity curves of the \hd\ system phased with the final orbit solution. The filled circles, open circles and triangles are the measured radial velocities of the components Ba, Bb and  A, respectively. The solid and dashed lines show the orbital solution of components Ba and Bb in the close binary. The phase is based on the ephemeris derived by the ORBITX solution.}
 \label{fig-rvcurves}
\end{figure}

Table\,\ref{tab-orbit} gives the orbital elements for the close system obtained by KOREL and ORBITX, respectively. The epoch T refers to the time of the maximum positive velocity of component Ba. 
In the ORBITX analysis, the eccentricity was found to be insignificant and was therefore fixed to zero, whereas in the KOREL analysis the eccentricity is small, but significant. Fig.\,\ref{fig-rvcurves} shows the orbit 
solution obtained by ORBITX for the narrow-line pair (circles) and the velocity of the broad-line component (triangles) phased with the ephemeris (HJD = 2\,453\,069.089(3) + 3.5744(3)$E$) of the close orbit.

\begin{table}
 \caption{Results for the orbit solution of the close binary.}
 \label{tab-orbit}
 \centering
 \begin{tabular}{l c c }
    
    \hline \hline
 & KOREL & ORBITX \\
    \hline

    P [d] & 3.5720 $\pm$ 0.0015 & 3.57436 $\pm$ 0.00034\\
    T (2\,453\,000+)& 69.1629 $\pm$ 0.0028 & 69.0889 $\pm$ 0.0025 \\
    e & 0.01003 $\pm$ 0.00036 & 0 (fixed) \\
    $\omega$ [\deg] & 7.59 $\pm$ 0.12 & - \\
    K$_1$ [\kms] & 74.564 $\pm$ 0.038 & 74.55 $\pm$ 0.24\\
    K$_2$ [\kms] & 77.00  $\pm$ 0.18 & 77.05 $\pm$ 0.27\\
    V$_0$ [\kms] & - & 38.32 $\pm$ 0.12\\
        
    \hline
  \end{tabular}
 
\end{table}

For \hd\ Ba and Bb, lower limits on the masses can be derived via the equation given by \citet{Aitken35}
\\
\\
$m_{1,2}\sin^3i = 10^{-6.98358}(K_1+K_2)^2K_{2,1}P(1-e^2)^{3/2}$ 
\\
\\
where $m$ denotes the mass in solar units, $K_1$ and $K_2$ represent the semi-amplitudes of the radial velocities in \kms. The period in days is denoted by $P$ and the eccentricity is expressed by $e$. The exponent has not been rounded off intentionally, to keep the formula true to original. The lower limits on the masses are: for component Ba, 0.68\,{\Msun}, and for component Bb, 0.65\,{\Msun}, which, together with similar luminosities deduced from line strength, suggests similar values of \Tf\ and \lg\ for both components.

Aitken's formula for the masses of binaries allows us to constrain the inclination angle of the close binary orbit. A conservative mass range for F type stars ranging from 1.1 to 1.6\,\Msun\ yields a range for the inclination angle of the orbital plane  between 47 and 57 degrees.

\subsection{
Determination of the Fundamental Parameters}
\label{subsect-fpar}

\subsubsection{Direct Method}
Further characteristics extracted from the LSD profiles are the mean radial velocity for component  A of $45.8\,\pm\,1\,$\kms, while the $\gamma$ velocity of the close binary Ba and Bb was confirmed to be 38.2\,$\pm$\,1\,\kms. A mean \vsini\ was derived for Ba to be $15.1\,\pm 1.1\,${\kms}; for Bb, $15.2\,\pm\,1.2\,${\kms}; and for  A,  $131\,\pm\,1.3\,${\kms}.  The relative equivalent width contributions are 
10\,$\pm$\,1\% for  Ba, 9\,$\pm$\,1\% for Bb, and 81\,$\pm$\,3\% for  A.
The analysis of the \Ha\ region was our main method of finding fundamental parameters of the components. 
The observed spectrum was fitted by scaling the individual model spectra using a parameter $a$, defined as follows:
\\ \\ 
$a = R_{ A}^{2}, R_{Ba}^{2} = R_{Bb}^{2} = \frac{1-a}{2}$, where $R_{Ba}^{2} + R_{Bb}^{2} + R_{ A}^{2} = 1$. 
\\ \\
The value of $a$ was adjusted simultaneously with effective temperatures. We arrived at a final value of $a$ = 0.7.
If the system constitutes a physical triple system, the parameter $a$ gives the radii ratio 
$R_{ A}$/$R_{ Ba,Bb} = \sqrt{0.7/0.15} = 2.2.$

\begin{figure}
\sidecaption
\resizebox{\hsize}{!} {
 \includegraphics*{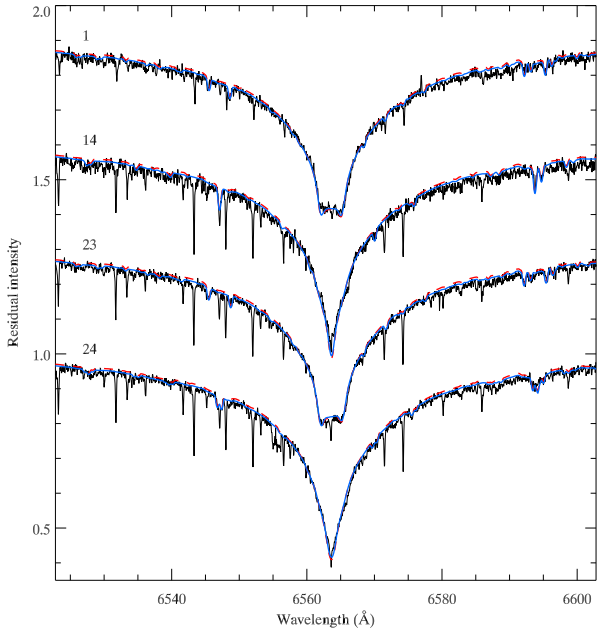}   
 }
 \caption{Fits to the H$\alpha$ line of the integrated spectrum of {\hd}. The black lines represent the observations; the solid line corresponds to a synthetic spectrum with \Tf\,=\,8000\,K, \lg\,=\,4.2, plus two stars with 6800\,K (4.4) and 6600\,K (4.4). Another model (dashed line) with 8000\,K (4.0), plus two stars with each 6000\,K (4.2), also matches the observed spectrum. Varying narrow lines are telluric in origin.}
 \label{fig-ha-fit}
\end{figure}

\begin{figure*}
\sidecaption
   \includegraphics[angle=90,width=13cm]{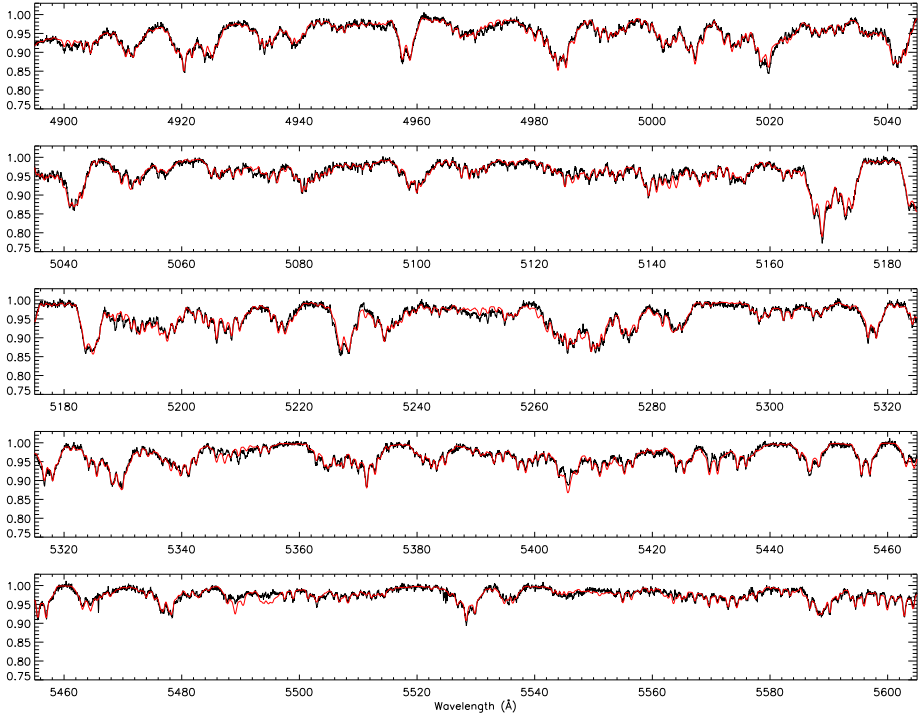}
 \caption{Comparison of our best-fitting synthetic spectrum to an observed spectrum of \hd\ with high S/N during a phase of large separation of components in the close binary, in a wavelength range with many metal lines. The model parameters for the synthesis are \Tf\,=\, 8000\,K, (\lg\,=\,4.2), 6800\,K (4.4) and 6600\,K (4.4), for A, Ba, Bb respectively.}
 \label{fig-metal}
\end{figure*}

\begin{figure}
 \resizebox{\hsize}{!} {
  \includegraphics*[angle=90]{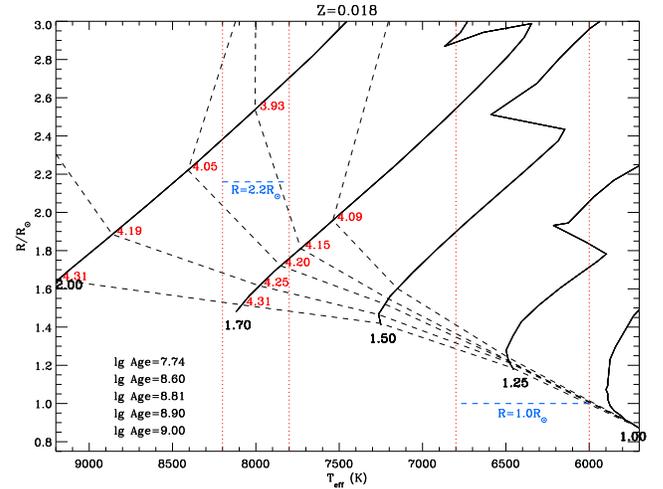}
 }
 \caption{Evolutionary tracks from \citet{Schaller} for masses \Msun\ $= 1.25 - 2.0$ and various \lg\ values, in the ranges we infer for \hd\ based on the spectral fits. The vertical dotted lines give the temperatures ranges for the individual components of \hd. The dashed lines correspond to isochrones with the given ages at the lower left corner. The isochrone for log age of 7.74 corresponds to the lowest line. }
 \label{fig-tracks}
\end{figure}

By comparing fits of synthetic triple-star spectra across a wide parameter space, we estimate for component  A an effective temperature of \Tf\ = (8000\,$\pm$\,200)\,K and for each component of the close binary \Tf\,=\,(6400$\,\pm$\,400)\,K. The rather large uncertainties of the temperatures for the close binary is caused by the uncertainty in \lg\ and the fact that its contribution to the total light is small compared to the dominant  A component. Fig.\,\ref{fig-ha-fit} illustrates the quality of the best fit and the accuracy to which fundamental 
parameters of the three components can be determined from existing data. The black wiggly lines represent four out of 29 observed spectra in the range around H$\alpha$ and the smooth blue line corresponds to a synthetic spectrum with \Tf\,=\,8000\,K and \lg\,=\,4.2,  plus two stars with 6800\,K (4.4) and 6600\,K (4.4). Another model with 8000\,K (4.0) plus two stars (each with \Tf\,=\,6000\,K  and \lg\,=\,4.2) also agrees with the observed spectra within the resolution. The radial velocity values for these fits were taken from our orbital solution (see Table 5).
The chosen parameters for the triple system give also a good fit to the metal line spectrum as is illustrated in Fig.\,\ref{fig-metal}.
Based on evolutionary tracks published by \citet{Schaller}, Fig.\,\ref{fig-tracks} shows that a triple system with log(age) between 8.8 and 8.9 is compatible with our observations of {\hd}.

\subsubsection{Disentangled Spectra}
The determination of the fundamental parameters was also performed on the disentangled spectra.
A grid of model spectra ranging from 6000 to 8600\,K with stepwidth of 200 K and log g of 3.6, 4.0, 4.4, respectively, was compared to the disentangled spectra corrected for radial velocity shifts. The model spectra were calculated using the code SynthV \citep{Tsymbal96} and NEMO model atmosphere grid \citep{Heiter2002}.
For finding the best fitting spectra two approaches have been used. 
First, the standard deviation of the residuals, hereafter referred to as $\sigma$, and second, the $\chi^2$ test were used for evaluating the quality of the fit.
These approaches lead to different best fitting solutions, because the methods differ in weighting the data. The standard deviation of the residuals weights each data point equally, while the $\chi^2$ test weights the data according to the relative line strength. Hence the $\chi^2$ test prefers better fits of line cores. Solar chemical composition was assumed, based on the good fit obtained by the direct method (see Fig.\,\ref{fig-metal}). 

\begin{figure}
 \resizebox{\hsize}{!} {
  \includegraphics*[]{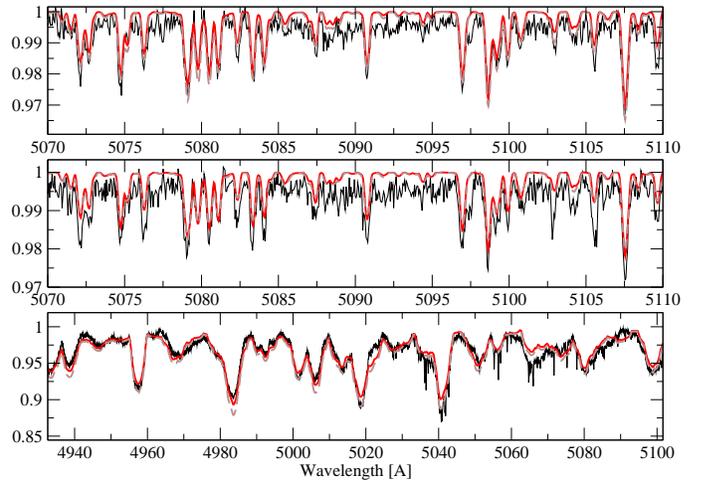}
 }
 \caption{Model spectra compared to the disentangled spectra. From top to bottom the comparison is shown for component  Ba, Bb and  A respectively. The full line corresponds to the best fitting model spectrum using $\sigma$, the gray dashed line corresponds to the results based on $\chi^2$ analysis. Details on the fitted model spectra are given in Table \ref{tab-korelparams}. Notice, that the scaling of the wavelength axis is different for  A and another wavelength region is shown.}
 \label{fig-korelcomp}
\end{figure}

The standard deviation of the residuals and $\chi^2$ was calculated the following way:
\\
\\
$\sigma = \sqrt{\frac{1}{N-1} \sum (R_i - \overline{R})^2 }$
\\ \\
$\chi^2 = \sum \frac{(S_{scal} - D)^2}{S_{scal}}$
\\
\\
where $S_{scal}$ means scaled synthetic spectrum and $D$ the disentangled spectrum and $R = S_{scal} - D$. 
The model spectra were scaled according to
\\ \\
\begin{large}
$S_{scal} = \frac{S f+1}{f+1}$
\end{large}
\\ \\
where $S$ denotes the model spectrum and $f$ corresponds to the luminosity factor.
This factor is the ratio of the luminosity of a given component to the sum of the luminosities of 
both other components.
For each model spectrum the best fitting luminosity factor $f$ and the optimal \vsini\ has been determined. 
There are spectral regions where the agreement of synthetic and disentangled spectra is rather poor, which may be due to an incomplete input line list. Those regions have been omitted for finding the final solution. Table\,\ref{tab-korelparams}
gives the best fitting model spectra for each component, which are illustrated in Fig.\,\ref{fig-korelcomp}.
As a conclusion, fitting model spectra to the disentangled spectra gives about the same result as the direct method discussed above. Both fainter components are cooler stars, while the brighter star is definitely hotter. The model spectra for the cooler stars are only weakly sensitive to changes in \lg, while the for the hotter component model spectra with lower \lg\ values give better $\sigma$ or $\chi^2$ values.

\begin{table}
\caption{Results for the individual spectra obtained by KOREL.}
\begin{tabular}{lllll}
\hline\hline
Component, method & T$_{\rm{eff}}$ & v sini & f & log g \\ 
\hline
 A, $\sigma$ & 8200 & 125 & 3.7 & 3.6 \\ 
 A, $\chi^2$ & 8000 & 131 & 8.5 & 3.6 \\
\hline
Bb, $\sigma$ & 6400 & 13 & 0.04 & 4.4 \\
Bb, $\chi^2$ & 6000 & 16 & 0.036 & 4.4 \\
\hline
 Ba, $\sigma$ & 6400 & 13 & 0.056 & 4.4 \\
 Ba, $\chi^2$ & 6000 & 15 & 0.06 & 4.4 \\
\hline

 \end{tabular}
 \label{tab-korelparams}
 \end{table}

Table\,\ref{tab-finalresults} shows the results of both analyses, disentangled spectra and H$_{\alpha}$ fitting of the combined spectrum. The most likely parameters are printed bold face. For the fainter components large uncertainties in the parameters are evident.

\section{Asteroseismic modelling of \hd\  A}
\label{sect-Model}

Theoretical pulsation frequencies of \hd\ were computed with updated codes for stellar evolution \citep{Paczynski69} and oscillation \citep{Dziembowski77}, where rotation is taken into account to second order \citep{Dziembowski92}. Standard Mixing Length Theory (MLT) for convection \citep{Boehm-Vitense58}  was used and the simplest convection - pulsation treatment (the assumption of frozen convective flux). Since envelope convection becomes less efficient for stars closer to the hot border of the classical instability 
strip, $\alpha_{MLT}$ can be expected to be smaller for \hd\ than for the \ds\ star FG Vir. For that star, the MLT parameter $\alpha_{MLT}$ was found to be 0.5 \citep{Daszynska-Daszkiewicz05}. 

Direct mode identification based on our photometric and spectroscopic data is not possible, so we instead investigate period ratios. The values of radial-pulsation period ratios depend mainly on metallicity and rotation (Suarez et al. 2006) and also slightly on the evolutionary stage of the star. Table\,\ref{tab-modelratios} shows the predicted values for radial period ratios of a \ds\ star model with Z\,=\,0.018 and 0.02 and V$_{\rm{rot}}$(ZAMS)\,=\,150\,\kms\ and 180\,\kms.

\begin{table}
\caption{Expected period ratios from models. }

\begin{tabular}{cccc}
\hline\hline
  & v$_{\rm{rot}}$ = 150\,\kms & v$_{\rm{rot}}$ = 150\,\kms  & v$_{\rm{rot}}$ = 180\,\kms\\
  & Z = 0.018                  &    Z = 0.02                  & Z = 0.018 \\
  \hline
  F/1H  & 0.782 & 0.782 & 0.786 \\
  1H/2H & 0.812 & 0.812 & 0.813 \\
  2H/3H & 0.842 & 0.841 & 0.843 \\
  3H/4H & 0.863 & 0.862 & 0.864 \\
  4H/5H & 0.880 & 0.880 & 0.882 \\
  5H/6H & 0.892 & 0.892 & 0.893 \\
\end{tabular}
\label{tab-modelratios}
\end{table}

For convenience, the radial fundamental mode is hereafter referred to as F, the first overtone as 1H, and so on. The observed pulsation {\freqs} in Table\,\ref{tab-freqs} were examined to identify possible period ratios for radial modes in {\hd}. (The two long periods given in Table\,\ref{tab-freqs} have been excluded because they are not due to \ds\ pulsation.)  Our temperature estimate indicates that the star is located close to the hot border of the instability strip of radially pulsating \ds\ stars \citep{Pamyat}. Under the assumption that radial modes are observed, we find five possibilities: 
\\
(a) $f_4$ / $f_5$ (1H/2H) = 0.819 \\
(b) $f_8$ / $f_{11}$ (2H/3H) = 0.837 \\
(c) $f_{12}$ / $f_{5}$ (2H/3H) = 0.846 \\
(d) $f_2$ / $f_{10}$ (2H/3H) = 0.841 \\
(e) $f_{11}$ / $f_{1}$ (4H/5H) = 0.884 \\

\begin{figure}
 \resizebox{\hsize}{!} {
   \includegraphics*{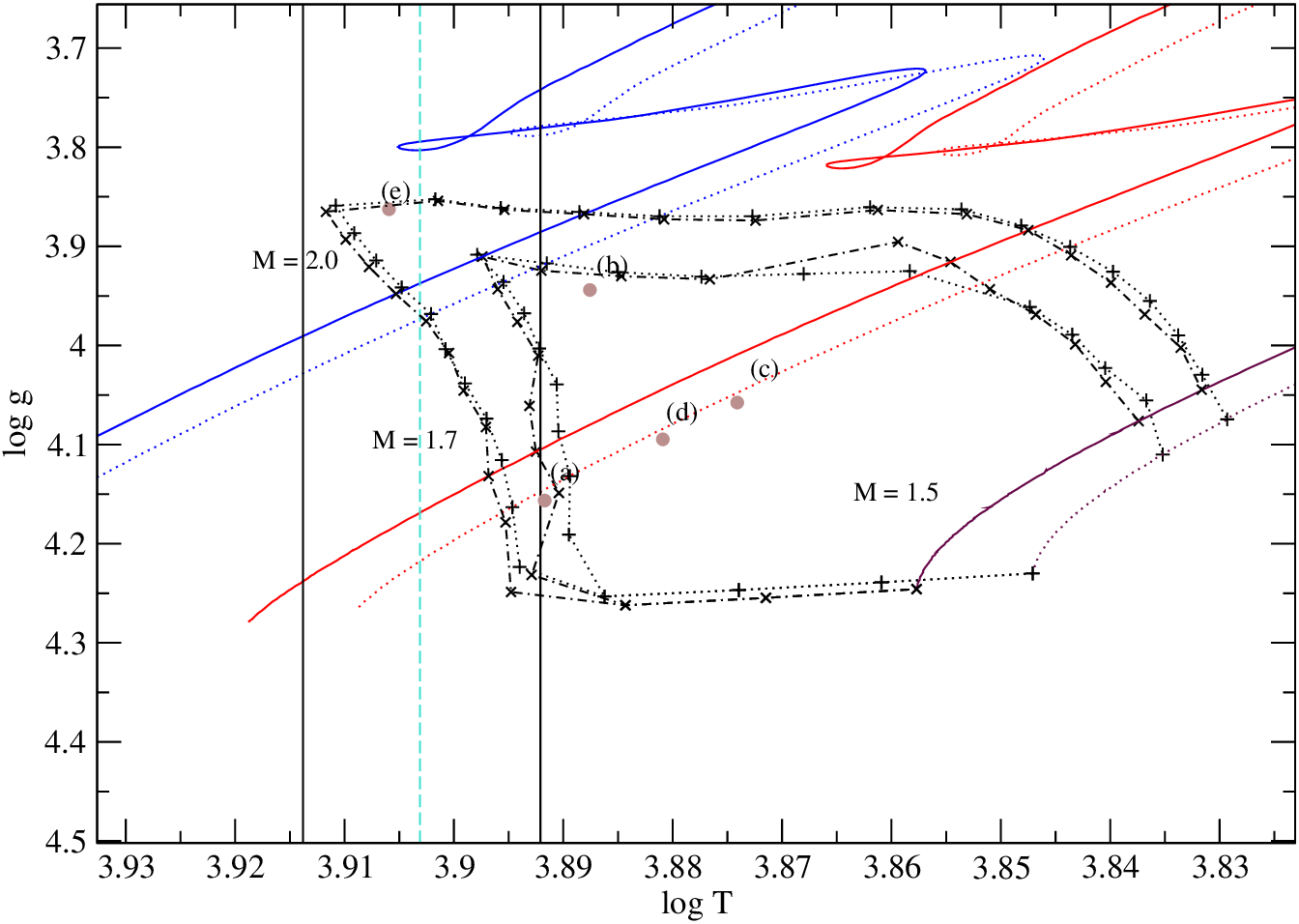}
 }
 \caption{Region in the \lg\ - log \Tf\ diagram where models fit the observed range of unstable mode frequencies in {\hd}. Evolutionary tracks for M\,=\,1.5, 1.75 and 2.0\,\Msun\ are given, where the full lines correspond to Z\,=\,0.018 and the dotted lines to Z\,=\,0.02. The vertical dashed line indicates the estimated effective temperature of component  A, and the error limits of the temperature of \hd\,A derived in Sec.\,\ref{subsect-fpar} are indicated by the full vertical lines. The filled circles labelled (a) to (e) are models which fit possible observed period ratios for radial modes (see text). The inner polygon shows the region of models where the observed range of unstable frequencies is exactly reproduced. The outer polygon shows the region where a deviation of 3 d$^{-1}$ is allowed. Both polygons have been calculated with Z = 0.018 (marked by the dashed-dotted line, crosses indicate the position of calculated models) and Z = 0.02 (marked by dotted line and pluses). }
 \label{fig-insrange}
\end{figure}

In fast rotating stars, nonradial modes usually dominate \citep{Breger99}, and this may also be the case for {\hd}. Unfortunately, fitting the observed frequencies to model frequencies with a least-squares technique does not result in a unique solution, because models predict a very dense spectrum of frequencies of which only a few (in this case, 11) are detected.  Hence we concentrated our modelling efforts on explaining the observed frequency {\em{range}} of $18.69 - 29.70$\,{\cd}.  Figure\,\ref{fig-insrange} shows the region in the \lg\,-\,log\,\Tf\ 
diagram where pulsation models are consistent with the frequency range observed in {\hd}. 
The inner polygons are calculated with Z\,=\,0.018 and Z\,=\,0.02 and represent the area in which all observed frequencies lie in the theoretically predicted range of excited modes. Due to uncertainties in stellar modelling and potentially incomplete opacity data the frequency range of unstable modes may be larger than predicted by our codes. The outer polygons shows the range of models where the observed frequency range is 3 \cd\ larger than the theoretically predicted range of unstable modes.

The models (a) to (e) mentioned above are also included in the figure. Models (b), (c) and (d) lie outside the errorbox, model (a) is close to the edge of the
error box, while (e) is inside. In model (e) the period ratio of the fourth overtone and the fifth overtone of the radial fundamental would match the theoretical ratio given in Table\,\ref{tab-modelratios}. However, the unstable modes range from 19.6\,\cd\ to 27.3\,\cd, which is too small by 1\,\cd\ on the lower frequency border and 2.3\,\cd on the higher frequency border, respectively. The model parameters are \Tf\,=\,8050\,K, \lg\,=\,3.86, M = 2.1 \Msun. This model 
is compatible to the derived parameters for \hd\ in Sect.\,\ref{sect-spectro}.  The amplitude of the presumably radial mode in model (e), $f_{11}$, is 0.16 mmag and did not change significantly, while the amplitude of the presumably consecutive radial mode, $f_{1}$, is about 1.5 mmag and was clearly variable. Considering  that $f_{11}$ is close to the detection limit while $f_{1}$ has the highest amplitude, it is questionable, if the identification with radial modes is correct in this case. 

The error box in  Fig.\,\ref{fig-tracks} is based on the assumption, that \hd\ forms a physical triple system, where for component  A \lg\,=\,4.0 was adopted.
Models hotter than $\log\,$\Tf$\,=\,3.92$ cannot explain the observed frequency range, nor can models with masses larger than 2.1 {\Msun}. 
Note, however, that models generated for FG Vir tend to be too cool by 200 to 300\,K, when comparing the best-fitting models to the spectroscopically determined temperatures.

\section{Discussion and Conclusions}
\label{sect-conc}

\begin{table*}
\centering
\caption{Summary of fundamental parameters derived with different techniques and proposed solution.}
\begin{tabular}{ l rrr rrr rrr}
\hline \hline
Component & \multicolumn{3}{c}{A} &  \multicolumn{3}{c}{Bb}  & \multicolumn{3}{c}{Ba} \\
Method       &  \Tf   & \lg    &  \vsini   & \Tf    & \lg    & \vsini      & \Tf    & \lg    & \vsini \\
\hline

KOREL ($\sigma$)& 8200  & 3.6   & 125 & 6400  & 4.4   & 13  & 6400  & 4.4  & 13 \\
KOREL ($\chi^2$)& 8000  & 3.6   & 131 & 6000  & 4.4   & 16  & 6000  & 4.4  & 15 \\
H$\alpha$          & 8000  & 4.0   & 131 & 6600  & 4.4   & 15  & 6800  & 4.4  & 15 \\
                        & 8000  & 4.2   & 131 & 6000  & 4.2   & 15  & 6000  & 4.2  & 15 \\

Seismology$^1$ max & 8100  & 4.2 &        & \\
                          min & 7500  & 3.85 &        & \\
Triple system$^2$ max &    & 4.3  &        & & 4.4    &     &        & 4.4   & \\
                            min &    & 3.9  &        & & 4.3    &     &        & 4.3   & \\
\hline
Proposed solution  & 8000  & 4.0 & 130 & 6400 & 4.4  &  15  & 6400 &  4.4  & 15  \\
estimated uncertainty    & 200   & 0.5   &   3 &  400 &  0.5   &   2  &  400 &  0.5    &  2  \\
\hline


\multicolumn{10}{l}{\footnotesize{$^1$ from frequency ranges predicted by theory, see Fig.\ref{fig-insrange},  $^2$ for chosen temperatures, see Fig.\,\ref{fig-tracks}.}}
\end{tabular}
\label{tab-finalresults}
\end{table*}

\hd\ was discovered to be a multiperiodic \dsct\ pulsator from MOST photometry in 2004.  Analysis of these data, and MOST data 
collected in 2005 and 2007, reveal 11 pulsation frequencies, with amplitudes as low as 90 ppm in the total flux of the system.  Subsequent Tautenburg spectroscopy demonstrates that \hd\ is a triple system, where the two form a close binary with a period of 3.57 days, and the \dsct\ star (which we dub the  A component and which contributes 80\% of the total flux) is either in a much wider orbit or is incidentally in the line of sight. The system does not exhibit 
eclipses, but variations with the period of the close binary orbit are evident in the 2007 MOST photometry. 
Another period of about 4 days is present in all three epochs of photometry. 
The origin of this period remains unknown.
The high \vsini\ of the $\delta$ Scuti star, and its likely radius, mean
that its rotation period must be about 1 day. The only way we know to
associate a 4-day period with the $\delta$ Scuti star is through an
isolated high-overtone g-mode. It could be a hybrid pulsator, with both
$\gamma$ Doradus g-modes and $\delta$ Scuti p-modes. However, the expected
range of high order g-mode periods is about 0.3 - 3 d, and most $\gamma$ Dor periods
are closer to 0.8 d. Thus the hybrid explanation is unlikely, and we conclude
the 4-day period originates in one or both components of the close binary.
However, the stars in the close binary are almost certainly tidally locked,
so it is unlikely that the 4-day period represents evidence for rotation in that system. 
If the rotation axes of the two components were
significantly misaligned, we would be able to measure the difference in
\vsini\ between the two stars in the spectra, and we do not see such a
difference.

An additional puzzle in the system is the fact that the orbital frequency of the close binary is equal to the difference between the two $\delta$ Scuti frequencies with the highest amplitudes. The separation of the $\delta$ Scuti star from the close binary is large enough that there should be no significant tidal influence. The observed frequency coincidence is remarkable, but we have no explanation for it.

Considering the various estimates for the fundamental astrophysical 
parameters of the three components we conclude that (\Tf/\lg) A: 
8000/4.0, Ba and Bb: 6400/4.4, are the best compromise within the uncertainties 
given in Table \ref{tab-finalresults}. This conclusion is also supported by the fact that  
synthetic spectra of a triple system computed for various combinations  
in the mentioned parameter space overlap within the observational 
uncertainties of our spectroscopy. The H$_{\alpha}$ line profiles are rather insensitive to \lg\ for stars with \Tf $<$ 8500\,K, hence the large uncertainty for \lg\ given in the table are estimated from the metal lines of the disentangled spectra.

The parallax given by \citet{Gatewood} is consistent with the A component being a \ds\ star with an absolute magnitude of  $M_V\,\approx\,1.8$.  For both fainter components, Ba and Bb, we derive $M_V\,\approx\,4.1$ and $4.2$, respectively, based on our model of the triple system.
\ds\ models enable us to estimate upper limits on the effective temperature (about $8000\,\rm{K}$) and mass (2.1 {\Msun}) of \hd\, A. Based on the orbit solution, we set lower mass limits of about 0.7 {\Msun} for the close binary components, Ba and Bb. 
However, given the expected masses for F-type stars (between 1.1 and 1.6 $M_{\odot}$), we can exploit the upper limit to constrain the inclination of the orbital plane to $47^{\circ} - 57^{\circ}$.

\begin{acknowledgements}
This project was supported by the Austrian Fonds zur F\"orderung der wissenschaftlichen Forschung (FWF, project \emph{The Core of the HR diagram}, P17580-N02), and the Bundesministerium f\"ur Verkehr, Innovation und Technologie (BM.VIT) via the Austrian Agentur f\"ur Luft- und Raumfahrt (FFG-ALR). VT also kindly acknowledges the Ukrainian grant FRSF F25.2/074 for partial financial support.  JMM, JFR, SR, DBG, and AFJM received research support from NSERC (Natural Sciences \& Engineering Research Council) Canada.  RK was partly funded by the Canadian Space Agency. 

\end{acknowledgements}

\bibliographystyle{aa}

\bibliography{9996bib.bib}

\end{document}